\documentclass[aip,amsmath,amssymb,reprint,longbibliography]{revtex4-2}

\usepackage{graphicx}
\usepackage{dcolumn}
\usepackage{bm}
\usepackage[utf8]{inputenc}
\usepackage[T1]{fontenc}
\usepackage{mathptmx}
\usepackage{upgreek}
\usepackage{physics}
\usepackage[colorlinks,plainpages=false,linkcolor=blue,urlcolor=blue,citecolor=blue,pdfpagemode=UseNone,pdfstartview=FitH]{hyperref}

\begin{document}

\title[Auger and spin dynamics in a self-assembled quantum dot]{Auger and spin dynamics in a self-assembled quantum dot}

\author{H. Mannel}
\email{hendrik.mannel@uni-due.de}

\author{J. Kerski}

\author{P. Lochner}
\author{M. Zöllner}
\affiliation{Faculty of Physics and CENIDE, University of Duisburg-Essen, 47057 Duisburg, Germany}

\author{A. D. Wieck}

\author{A. Ludwig}
\affiliation{Lehrstuhl für Angewandte Festk\"orperphysik, Ruhr-Universit\"at Bochum, 44780 Bochum, Germany}

\author{A. Lorke}

\author{M. Geller}
\affiliation{Faculty of Physics and CENIDE, University of Duisburg-Essen, 47057 Duisburg, Germany}

\date{\today}

\begin{abstract}
	The Zeeman-split spin states of a single quantum dot can be used together with its optical trion transitions to form a spin-photon interface between a stationary (the spin) and a flying (the photon) quantum bit. Besides long coherence times of the spin state itself, the limiting decoherence mechanisms of the trion states are of central importance. We investigate here in time-resolved resonance fluorescence the electron and trion dynamics in a single self-assembled quantum dot in an applied magnetic field of up to $B = 10\,$T. The quantum dot is only weakly coupled to an electron reservoir with tunneling rates of about $1\,$ms$^{-1}$. Using this sample structure, we can measure, in addition to  the spin-flip rate of the electron and the spin-flip Raman rate of the trion transition, the Auger recombination process, that scatters an Auger electron into the conduction band. The Auger effect destroys the radiative trion transition and leaves the quantum dot empty until an electron tunnels from the reservoir into the dot. The Auger recombination rate decreases by a factor of three from $\gamma_A=3\,\mu$s$^{-1}$ down to $1\,\mu$s$^{-1}$ in an applied magnetic field of $10\,$T in Faraday geometry. The combination of an Auger recombination event with subsequent electron tunneling from the reservoir can flip the electron spin and thus constitutes a previously unaccounted mechanism that limits spin coherence, an important resource for quantum technologies.
\end{abstract}

\maketitle

\section{Introduction}

Self-assembled quantum dots (QDs)\cite{Bimberg1999,Petroff2001} are promising nanostructures that can host a single spin\cite{Atature2006,Imamogextasciimacronlu1999,Hanson2008,Benjamin2009} to realize a quantum bit (qubit) in a solid-state environment.\cite{Ladd2010} The Zeeman-split states of the electron/hole spin in a magnetic field form here the two-level system for such a stationary qubit.\cite{Calarco2003} This spin qubit can be controlled via fast optical pulses,\cite{Press2008,Gerardot2008} using the optically accessible trion states consisting of a pair of electrons in a spin-singlet state and an unpaired heavy-hole with spin-up or spin-down. However, to build a quantum computer,\cite{Ladd2010} or even more, a quantum internet,\cite{Kimble2008} distant spins have to be transferred and entangled.\cite{Delteil2015, Bernien2013, Hensen2015} A prerequisite for this is that the coherent quantum state of the spin has to be mapped onto a flying qubit, a single photon\cite{Atatuere2018} in a spin-photon interface.\cite{Javadi2018, Gao2012} Hence, long spin dephasing times for the stationary qubit and long coherence times for highly indistinguishable photons are needed for these envisioned quantum information technologies.

Single photons from (InGa)As self-assembled QDs have shown long coherence times,\cite{Matthiesen_2012} high indistinguishability,\cite{Matthiesen2013, Santori2002} and long spin lifetimes.\cite{Gillard2021} However, a commonly neglected process has come more into the focus, as it limits the coherence time of the spin and trion states: The Auger effect as an electron-electron scattering process,  where the recombination energy of a trion  is transferred to another excess electron/hole, that is ejected out of the QD into the conduction/valence band continuum. This scattering process is well known from colloidal QDs\cite{Vaxenburg2015, Cohn2013, Efros1997} and has been directly observed just recently in resonance fluorescence measurements\cite{Loebl2020} with Auger recombination rates in the order of microseconds.\cite{Geller2019, Beckel2014, Lochner2020, Kurzmann2016} 

In this paper, we show time-resolved resonance fluorescence (RF) measurements on the trion transitions of a negatively-charged InAs QD, embedded in an electrically controllable diode structure and charged by electron tunneling from a nearby charge reservoir. An applied magnetic field $B$ of up to $10\,$T in Faraday geometry (here, parallel to the growth axis) splits the trion into a lower (''red'') and a higher (''blue'') energy transition. We address one of the trion transitions with a tunable diode laser and measure time-resolved the decay of the fluorescence intensity. The observed transients involve the processes that quench the RF signal: The previously studied spin-flip between the up- and down-spin states of the electron,\cite{Atature2006, Lu2010, Kroutvar2004} the spin-flip Raman scattering\cite{Dreiser2008} and, moreover, the Auger recombination. Using a rate equation model and a fit to the transients allows us to obtain the evolution of these scattering rates as a function of the applied magnetic field. Interestingly, we observe a decreasing Auger rate from $3\,\mu\text{s}^{-1}$ down to $1\,\mu\text{s}^{-1}$ for increasing magnetic field. After an Auger recombination has emptied the QD, eventually an electron with opposite spin tunnels from the reservoir into the QD. This leads to an Auger-assisted spin dynamic which limits the spin lifetime in a self-assembled QD, for an optical spin read-out via the trion transition. A more detailed understanding of the Auger effect in self-assembled QDs will help to decrease the Auger recombination and enhance the spin-lifetime in future optimized QD structures. 

\section{Sample design and methods}

The measurements were performed on a single self-assembled (InGa)As QD at $4.2\,$K in a confocal microscope setup. The sample contains a single layer of QDs and was grown by molecular beam epitaxy.\cite{Petroff2001} During the growth process, the QDs were In-flushed\cite{Wasilewski1999} to shift the emission wavelength to $\approx950\,$nm. The QD layer is embedded in a p-i-n diode structure with a highly n-doped GaAs layer as a charge reservoir and a highly p-doped GaAs layer as an epitaxial gate\cite{Ludwig2017}. Between the charge reservoir and the QD, $45\,$ nm (AlGa)As are implemented as tunneling barrier to achieve electron tunneling times in the order of milliseconds  (see Lochner et al.\cite{Lochner2020} for details about the sample structure). An applied voltage between charge reservoir and gate can control the charge state of the QD.\cite{Hoegele2004,Geller2019} Furthermore, the resonance of the QD can be tuned by the quantum confined Stark effect.\cite{Li2000} The QD is optically investigated by resonance fluorescence, where the laser background is suppressed by cross polarization.\cite{Yilmaz2010}

\section{Quantum dot states and transitions in a magnetic field}

\begin{figure}
	\includegraphics[width=\columnwidth]{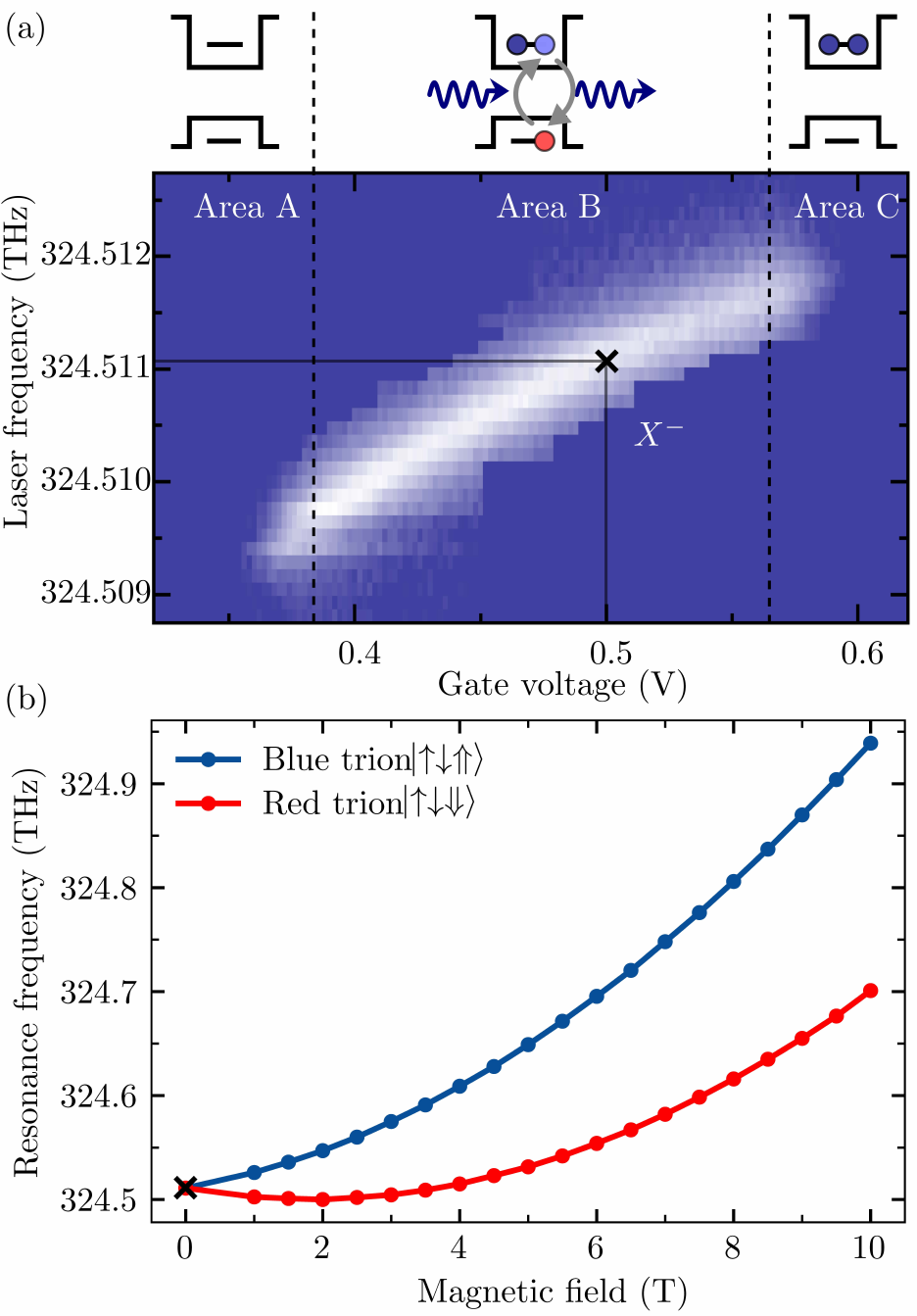}
	\caption{(a) RF intensity of the trion transition as a function of gate voltage and laser frequency. In the interval from $V_\mathrm{G}=0.38\,$V to $0.58\,$V (area B), the QD is charged with one electron and the trion transition is observed at laser frequencies between 324.5095\,THz and 324.5115\,THz. In area A, the QD is empty. In area C, it is charged with two electrons. (b) Magnetic field dependence of the trion resonance frequencies at $V_G=0.5\,$V. The data point at $B=0$ is taken from (a) (black cross). At $B\not=0$, the trion transition splits into an energetically lower (red) and an energetically higher (blue) transition.}
	\label{fig:trion} 
\end{figure}

In Fig.~\ref{fig:trion} (a), the RF signal of the negatively-charged exciton, the so-called trion (X$^-$), is shown as a function of gate voltage and laser frequency. In the gate voltage range from $0.38$ to $0.58\,$V (area B), the QD is charged with one electron, and RF from the trion can be observed. At lower gate voltages (area A), the QD is empty while at higher voltages (area C), it is charged with two electrons. In both cases, the trion transition is forbidden. 
When a magnetic field (Faraday geometry) is applied to the sample, the trion state will not be spin-degenerated anymore, and will split up into an energetically lower ("red trion" with spin configuration $\ket{\uparrow\downarrow\Downarrow}$) and an energetically higher state ("blue trion" with spin configuration $\ket{\uparrow\downarrow\Uparrow}$) .\cite{Atature2006} For the specific QD investigated here, this can be seen in Fig.~\ref{fig:trion} (b). In this measurement, a gate voltage of $0.5\,$V is applied, and the laser frequency is tuned to match the blue and red trion resonance. The energy difference between the two trion resonances increases linearly due to the Zeeman effect,\cite{ZEEMAN1897} while their mean value follow a diamagnetic shift.

\begin{figure}
	\includegraphics[width=\columnwidth]{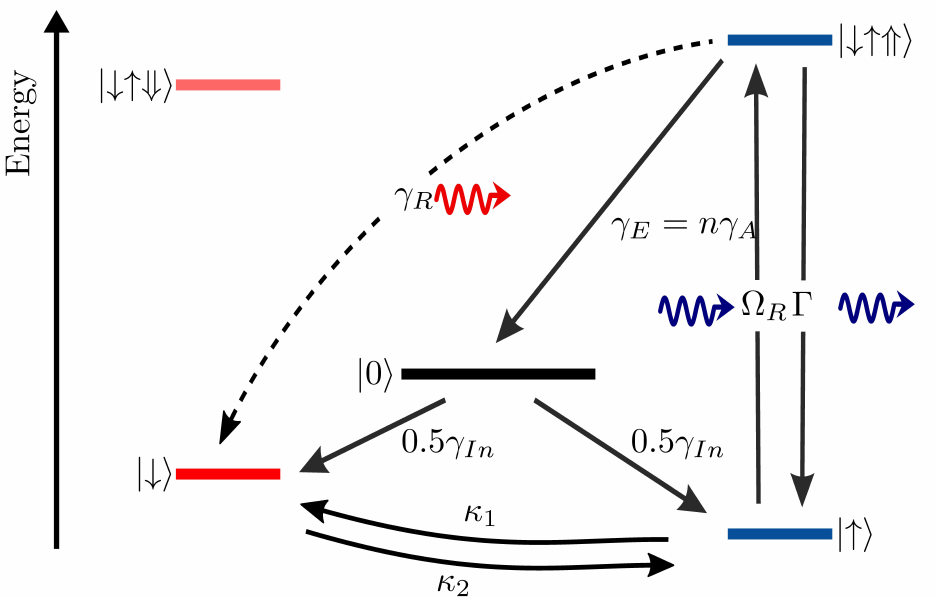}
	\caption{Level scheme of the singly-charged QD in a magnetic field (Faraday geometry) with transitions. The spin states $\ket{\downarrow}$ and $\ket{\uparrow}$ are energetically splitted by the Zeeman effect. The optical transition to the energetically higher blue trion state $\ket{\uparrow\downarrow\Uparrow}$ is excited with the Rabi frequency $\Omega_R$ and decays spontaneously with rate $\Gamma$. The trion state can also decay via a spin-flip Raman process with the rate $\gamma_R$ into the spin state $\ket{\downarrow}$ or via a non-radiative Auger recombination with emission rate $\gamma_E$ into the crystal ground state $\ket{0}$. By tunneling of an electron with random spin orientation, states $\ket{\downarrow}$ and $\ket{\uparrow}$ get occupied. The electron spin can flip from $\ket{\downarrow}$ to $\ket{\uparrow}$ with the spin relaxation rate $\kappa_2$ and inversely with rate $\kappa_1$. 
	}
	\label{fig:Ratenmodell} 
\end{figure}

Figure \ref{fig:Ratenmodell} schematically shows the energy diagram in a magnetic field with the spin-up $\ket{\uparrow}$ and spin-down $\ket{\downarrow}$ ground states that form, together with the transitions to the red ($\ket{\uparrow\downarrow\Downarrow}$) and blue trion ($\ket{\uparrow\downarrow\Uparrow}$), two Lambda-schemes.\cite{Dreiser2008,Kroner2008,Fernandez2009,Warburton2013} Here, we add the crystal ground state $\ket{0}$ to the Lambda-schemes in Fig.~\ref{fig:Ratenmodell}, as this state is accessible by the Auger recombination process (see discussion below). In a magnetic field, we resonantly excite the blue transition by a laser field with Rabi frequency $\Omega_R$ to the trion configuration  $\ket{\uparrow\downarrow\Uparrow}$, which is an electron singlet state with an additional heavy-hole with spin projection $m_z=3/2$ (see Fig.~\ref{fig:Ratenmodell}). An optical transition with the spontaneous emission rate $\Gamma$ emits a photon, and the spin-up ground state $\ket{\uparrow}$ is recovered. In addition, with much smaller probability (branching rations $\Gamma/\gamma_R$ of $10^3$ above $B=60$\, mT\cite{Dreiser2008}), a spin-flip Raman process\cite{Debus2014} with rate $\gamma_R$ scatters to the $\ket{\downarrow}$ state by a heavy-hole spin-flip and emission of a photon. Because of the energy mismatch between the ''blue'' laser excitation and the ''red'' trion transition, the RF signal is switched off until a spin relaxation into the initial spin-up state $\ket{\uparrow}$ with rate $\kappa_2$ has occurred. The spin-flip is mediated by spin-orbit coupling\cite{Kroutvar2004} or hyperfine interaction to the nuclei-spin bath.\cite{Khaetskii2002,Urbaszek2013,Kuhlmann2013} It is also possible that a spin-flip in the QD occurs in the opposite direction from  $\ket{\uparrow}$ to $\ket{\downarrow}$ with the rate $\kappa_1$. The ratio of $\kappa_1$ and $\kappa_2$ will be discussed later. As long as the QD is in the $\ket{\downarrow}$ state, no blue trion transition is possible. Thus, both, the spin-flip Raman process and the direct spin-flip from the $\ket{\uparrow}$-state to the $\ket{\downarrow}$-state, reduce the RF signal by scattering into the dark spin-down $\ket{\downarrow}$ state.  

Another important effect, which can switch off the RF of the blue trion transition, is the Auger effect (arrow from $\ket{\downarrow\uparrow\Uparrow}$ to $\ket{0}$ in Fig.~\ref{fig:Ratenmodell}). In this non-radiative process, the recombination energy of an electron-hole pair is transferred to the additional electron in the dot, which subsequently is emitted with rate $\gamma_E$ into the conduction band. Hence, the QD is in the crystal ground state $\ket{0}$. The Auger-mediated electron emission rate $\gamma_E$ is proportional to the Auger rate $\gamma_A$ and the average occupation of the QD with a trion $n$: $\gamma_E= n \cdot\gamma_A $.\cite{Kurzmann2016,Lochner2020} The Auger recombination is often neglected, especially in samples with small tunneling barriers and tunneling rates faster than the spontaneous emission lifetime, which is in the order of $\Gamma \approx 1\,\mathrm{ns}^{-1}$.\cite{Narvaez2005} In these samples with strong tunnel-coupling, the Auger-emitted electron is replaced immediately by tunneling from the charge reservoir. However, the Auger effect still decreases the emitted intensity, and randomizes the electron spin by Auger-assisted spin dynamics, as the QD can be recharged with an electron by tunneling from the reservoir with rate $\gamma_{In}$ either into the spin-up $\ket{\uparrow}$ or the spin-down state $\ket{\downarrow}$. From a dark spin-down state $\ket{\downarrow}$, the QD can be reset to the bright spin-up state $\ket{\uparrow}$ by a spin-flip with rate $\kappa_2$. A spin-flip in the opposite direction is also possible with the rate $\kappa_1$.\cite{Dreiser2008} In summary: As the blue trion can only be excited from state $\ket{\uparrow}$, there are three processes, which decrease the RF signal: (i) A direct spin-flip from the spin-up to the spin-down state with rate $\kappa_1$, (ii) a spin-flip Raman process with rate $\gamma_R$, or (iii) an Auger recombination with rate $\gamma_E$. For process (i) and (ii), a subsequent spin-flip with rate $\kappa_2$ is required to return to the optical bright state $\ket{\uparrow}$. The Auger process for case (iii) needs electron tunneling from the charge reservoir with rate $\gamma_{In}$ and, if the QD is recharged with a spin-down electron, an additional spin-flip to get to the bright spin-up state $\ket{\uparrow}$ again.

\begin{figure*}
	\includegraphics[width=\textwidth]{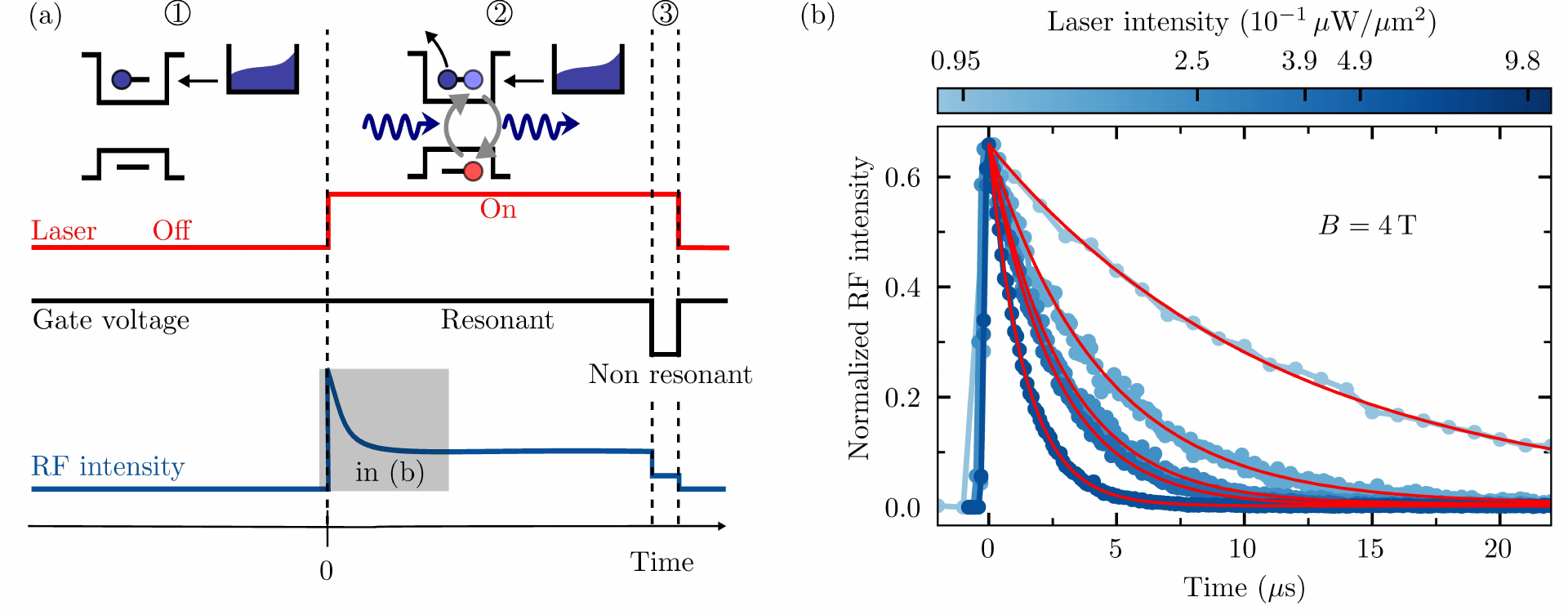}
	\caption{Time-resolved RF n-shot measurement of the Auger effect and spin dynamics in a magnetic field. (a) One shot consists of the preparation (1) and probing (2) of the QD and the background correction of the signal (3). (1) The QD is prepared with one electron by switching off the laser (red line) and setting $V_\mathrm{G}=0.5\,$V (black line). (2) The laser (resonant with the blue trion transition) is switched on, and $V_\mathrm{G}$ remains at $0.5\,$V. As the QD is prepared with one electron, the RF signal (blue line) is expected to switch on and to decrease afterwards due to Auger recombination and the spin-flip Raman process. (3) For the background correction of the measurement, the gate voltage is changed to $-0.5\,$V, the QD is off resonance and thus, only the laser background is measured to be subtracted. (b) Normalized RF intensity in area (2) for a magnetic field of 4 T. While the laser is turned on and the QD is in resonance, decreasing transients are observed. The measured transients (blue) along with fits by the rate model (red lines) are shown for different laser intensities.}
	\label{fig:Messung}
\end{figure*}

To investigate the described Auger and spin dynamics of charge carriers in a single self-assembled QD in a magnetic field, we use time-resolved RF. The measurement scheme can be seen in Fig.~\ref{fig:Messung}(a), where the laser intensity, the gate voltage, and the expected RF signal are displayed. A single shot consists of three parts: The preparation (1), the probing of the QD (2), and the background correction of the signal (3). In (1), the QD is prepared in the ground state. That means, one electron is charged in the QD, hence, it is either in state $\ket{\uparrow}$ or in state $\ket{\downarrow}$. To achieve this, the laser is turned off and the gate voltage is set to $0.5\,$V (which is in area~B in Fig.~\ref{fig:trion}(a)) for $2\,$ms, which is longer than the tunneling time $t_{In}=1/\gamma_{In}$ plus the time until a thermal equilibrium between states $\ket{\uparrow}$ and $\ket{\downarrow}$ is reached.

The occupation probability of these states in thermal equilibrium at the end of the preparation step (1) has a minor effect on the determination of the involved rates, however, it can be easily estimated: In equilibrium, the occupation of both states follow a Boltzmann distribution at $4.2\,$K\cite{Dreiser2008} with the energy splitting of the spin-up state $\ket{\uparrow}$ and the spin-down state $\ket{\downarrow}$. The energy of the spin-splitting is approximated by using the energy splitting of the red and blue trion resonance as shown in Fig.~\ref{fig:trion}(b). With an electron $g$-factor at our transition energy of 1.33 eV of about $g_e=0.8$\cite{Debus2014} and an hole $g$-factor of $g_h=0.2$ the hole is neglected here as a reasonable approximation. As an example, for a magnetic field of $4\,$T, this results in thermal occupation probabilities of about 65\,\% for the spin-up state $\ket{\uparrow}$ and 35\,\% for the spin-down state $\ket{\downarrow}$. The difference between the occupation probabilities increases up to an occupation probability of 93\,\% for $\ket{\uparrow}$ and 7\,\% for $\ket{\downarrow}$ at $B= 10\,$T. The ratio of occupation probabilities of states $\ket{\uparrow}$ and $\ket{\downarrow}$ will also be used to describe the ratio of the spin-flip rates $\kappa_2$ and $\kappa_1$.

In step (2) in Fig.~\ref{fig:Messung}(a), at $t=0$, the laser is switched on and excites resonantly the blue trion transition while the RF signal is recorded. As the excited blue trion can also decay via Auger-recombination or the spin-flip Raman process, the QD can end up either in the crystal ground state $\ket{0}$ or the spin-down state $\ket{\downarrow}$ (see Fig.~\ref{fig:Ratenmodell}); no RF signal can be recorded until an electron tunnels into the dot or a spin-flip has occurred. In an n-shot time-resolved measurement, a decreasing transient of the RF intensity is observed, starting with a maximum at $t=0$ (see blue line in Fig.~\ref{fig:Messung} (a)) and a saturation behavior for $t\gg0$ until dynamic equilibrium is reached. The trion transition is turned "off" in equilibrium by electron emission via Auger recombination or spin-flip Raman scattering and "on" again by electron tunneling or a spin-flip process (see Fig.~\ref{fig:Ratenmodell}). At step (3) in the time sequence, the gate voltage is switched to $-0.5\,$V (area A in Fig.~\ref{fig:trion}(a)) to measure the laser background and APD dark counts for background correction. For a good signal to noise ratio, $50,000$ to $100,000$ transient shots were accumulated.

These time-resolved n-shot RF measurements are performed for magnetic fields of 0, 4, 6, 8 and $10\,$T. For each magnetic field, the time-resolved RF signal is recorded at different laser excitation intensities ranging from $9.5\cdot10^{-2}\,\mu$W$/\mu$m$^{2}$ up to $9.8\cdot10^{-1}\,\mu$W$/\mu$m$^{2}$. As expected, the RF-signal shows decreasing transients starting from $t=0$. As an example, the measured transients at a magnetic field of $4\,$T can be seen in Fig.~\ref{fig:Messung}(b) (blue) for different laser intensities. The measured background is subtracted for each transient, and they are normalized to the occupation probability of the spin-up state $\ket{\uparrow}$, which is 65\,\% at $4\,$T (see discussion above). With increasing laser intensity, the RF intensity decreases faster. For all shown laser intensities, the transients saturate after $<0.06\,$ms.

\section{Results and Discussion}

In the following, we present a rate equation model to fit the obtained transients. The emission rate $\gamma_E$ for the Auger recombination from the blue trion state $\ket{\uparrow\downarrow\Uparrow}$ to the crystal ground state $\ket{0}$ is given by $\gamma_E= n \gamma_A$, where $n$ is the occupation probability of the QD with a blue trion and $\gamma_A$ the Auger rate.\cite{Kurzmann2016,Lochner2020} To obtain the occupation probability $n$, we measured for every magnetic field the RF intensity at $t=0$ as a function of the laser excitation intensity and fitted the data with a saturation function for a two level system\cite{Loudon2000}, see supplementary material for more details. The tunneling rate into the dot $\gamma_{In}$ is determined for every magnetic field in a separate measurement (see supplementary material).

With these parameters, three rate equations describe the probability to be in state $\ket{0}$, $\ket{\downarrow}$ or $\ket{1}$ for a given time $t$; where $\ket{1}$ represents being in the the process of optical driving the blue trion transition. The rate equations, which describe the system shown in Fig.~\ref{fig:Ratenmodell}, are 

\begin{align}
	\dot{P}_{\ket{0}}(t)&=n\cdot \gamma_A\cdot P_{\ket{1}}(t) - \gamma_{In}\cdot P_{\ket{0}}(t),\label{eq:1}\\
	\begin{split}
		\dot{P}_{\ket{\downarrow}}(t)&=0.5\cdot \gamma_{In}\cdot P_{\ket{0}}(t) + (1 - n) \cdot \kappa_1\cdot P_{\ket{1}}(t)\label{eq:2}\\& + n\cdot \gamma_R\cdot P_{\ket{1}}(t) - \kappa_2\cdot P_{\ket{\downarrow}}(t),
	\end{split}\\	
	\begin{split}
		\dot{P}_{\ket{1}}(t)&=0.5\cdot \gamma_{In}\cdot P_{\ket{0}}(t) + \kappa_2\cdot P_{\ket{\downarrow}}(t)\\& - (1 - n)\cdot \kappa_1 \cdot P_{\ket{1}}(t)\label{eq:3} - n\cdot (\gamma_A + \gamma_R)\cdot P_{\ket{1}}(t).
	\end{split}
\end{align}

\begin{figure}
	\includegraphics[width=\columnwidth]{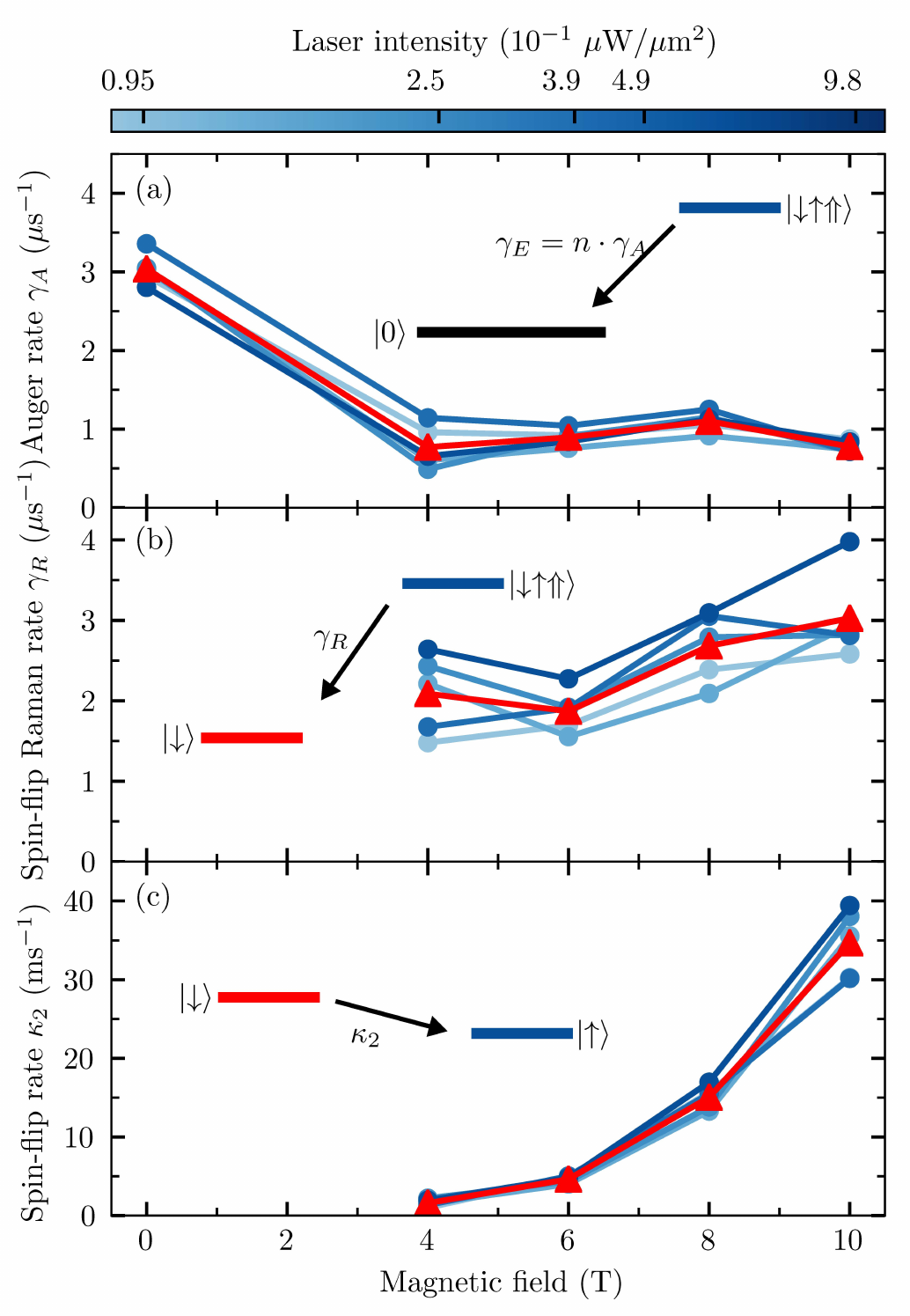}
	\caption{Magnetic field dependence of the Auger rate (a), spin-flip Raman rate (b) and spin-flip rate (c) for different laser intensities (blue shades), obtained from fits of the numerical solution of Eq.~(\ref{eq:3}). The average rates over all laser intensities for each magnetic field are shown in red. At $0\,$T, the trion transition is degenerate and we are unable to extract spin-flip nor spin-flip Raman processes with our method.} 
	\label{fig:ergebnis}
\end{figure}

The sum of all probabilities $P_{\ket{0}}$, $P_{\ket{\downarrow}}$ and $P_{\ket{1}}$  must be $1$ for every time $t$. The initial conditions at $t=0$, as well as the ratio of $\kappa_1$ and $\kappa_2$, are given by the earlier described Boltzmann distribution for each magnetic field. At $B=4\,$T, the conditions are $P_{\ket{0}}(0)=0$, $P_{\ket{\downarrow}}(0)=0.35$ and $P_{\ket{1}}(0)=0.65$. The measured RF intensity is proportional to the rate of photons emitted from the radiative trion recombination. $\dot{P}_{\ket{1}}(t)$ describes the temporal development of the normalized RF intensity and is numerically integrated with the three free parameters; the Auger rate $\gamma_A$, the spin-flip rate from state $\ket{\downarrow}$ to state $\ket{\uparrow}$ $\kappa_1$ and the spin-flip Raman rate $\gamma_R$. ${P}_{\ket{1}}(t)$ is used to fit the measured data for every excitation laser intensity in each magnetic field. For $4\,$T, the fits are shown as red lines in Fig.~\ref{fig:Messung}(b) in good agreement with the blue data points. From these fits we have determined the values of the Auger, spin-flip Raman and spin relaxation rates in Fig.~\ref{fig:ergebnis}. The blue lines in Fig.~\ref{fig:ergebnis} show the rates for increasing laser intensity from $0.095$ up to $0.98\,\mu $W/$\mu $m$^2$. We do not see a significant change in the involved rate for different laser excitation intensities. Thus we will concentrate on the average rates, shown as red lines in Fig.~\ref{fig:ergebnis}. We have checked the quality for three fitting parameters by fixing different rates at different values (see supplementary material for more information). Also, we have compared the spin-flip rate $\kappa_2$ and the spin-flip Raman rate $\gamma_R$ with values from the literature, discussed in the following.

The spin-flip rate, shown in Fig.~\ref{fig:ergebnis}(c), has been measured before, for instance by Kroutvar et al.\cite{Kroutvar2004} and Lu et al.\cite{Lu2010}, and gives, hence, a good checkpoint for our fitting routines for the unknown Auger and spin-flip Raman rate. We observe a spin-flip rate of $\kappa_2= 1.58$ ms$^{-1}$ at $B= 4 \,$T, increasing up to 34.8 ms$^{-1}$ at $B= 10 \,$T;  having a power-law dependence with an exponent of $m=3.4$ (see supplementary material). These values and the dependence of the spin-flip rates are in good agreement with those from the literature. The spin-flip Raman rate (Fig.~\ref{fig:ergebnis}(b)) tends to increase with increasing magnetic field from an average value of $2.0\,\mu\mathrm{s}^{-1}$ at $4\,$T to an average value of $3.0\,\mu\mathrm{s}^{-1}$ at $10\,$T, however, this increase is within the error margin of our measurements. We show here that the spin-flip Raman scattering rate $\gamma_R$ is up to 10 T always more than two order of magnitude faster than the spin-flip rate. This observation is in good agreement with the opportunity to use the spin-flip Raman scattering process for spin-pumping,\cite{Lu2010} since the Raman process scatters faster into the spin-down state $\ket{\downarrow}$ than the spin-flip can scatter back into the spin-up state $\ket{\uparrow}$. This is even more effective at smaller magnetic fields, where the ratio $\gamma_R / \kappa_2$ exceeds a factor of $10^3$. This ratio also explains the lack of data between 0\,T and 4\,T. The spin-up electron can be pumped into the dark state (here $\ket{\downarrow}$) via a spin-flip Raman process, while at the same time the spin-flip rate $\kappa_2$ is very small in low magnetic fields (see Fig.~\ref{fig:ergebnis}(c)). As a consequence, it takes a long time to return into the resonantly excited blue trion state and the RF signal is too small to be measured. At higher magnetic fields, the spin-flip rate is fast enough to recover the RF of the blue trion.

Finally, Fig.~\ref{fig:ergebnis}(a) shows the Auger rate $\gamma_A$ at $B=0$ and $B=4$ to $10\,$T. At $B=0$, the Auger rate is around 3\,$\mu\mathrm{s}^{-1}$ for all laser excitation intensities;  in agreement with previous measurements on almost identical InAs/GaAs QDs.\cite{Kurzmann2016,Lochner2020}  At $B=4\,$T, the Auger rate has dropped below 1\,$\mu\mathrm{s}^{-1}$ and is, within the accuracy of the measurement, almost constant up to $B=10\,$T. This is in contrast to a naive expectation of a increasing Auger rate with increasing magnetic field. Indeed, additional magnetic confinement is evidenced by the diamagnetic shift (see Fig.~\ref{fig:trion}(b)). This stronger carrier confinement implies a larger carrier overlap and thus should increase the Auger rate. We observe the opposite, so our finding needs another explanation. Further experimental studies and theoretical calculations including the final density of states for the Auger-scattered electron in the conduction band will shed light to our observation.

\section{Conclusion}

In conclusion, we have presented time-resolved resonance fluorescence measurements on the blue trion transition of a self-assembled QD in an applied magnetic field. The resonant pumping of the trion transition and the decay of the fluorescence intensity allows to study the underlying processes, where in a weakly tunnel-coupled dot (tunneling rates of about 1/ms) not only the spin-flip and spin-flip Raman processes but the Auger recombination can be observed. The spin-flip rate increases for increasing magnetic field. The spin-flip Raman rate $\gamma_R$ increase slightly in a magnetic field while the Auger rate $\gamma_A$ decrease by a factor of three. The combination of an Auger recombination and electron tunneling event has a 50\% chance to flip the electron spin. The Auger-rate is several orders of magnitude larger than the spin-flip rate. Thus, it can significantly limit the spin lifetime in quantum-dot-based devices for quantum information technologies, while longer spin lifetimes are desired. Besides adapting the QD shape and size, our study points towards an additional tuning knob: an Auger rate reduction seems to be possible by band-structure engineering of the final density of states of the emitted Auger electron in an advanced sample structure.

\vspace{1cm}
The data that support the findings of this study are available from the corresponding author upon reasonable request.

\begin{acknowledgments}
	This work was funded by the Deutsche Forschungsgemeinschaft (DFG, German Research Foundation) – Project-ID 278162697 – SFB 1242, and the individual research grant No. GE2141/5-1. A. Lu. acknowledge gratefully support of the DFG by project LU2051/1-1. A. Lu. and A. D. W. acknowledges support by DFG-TRR160, BMBF - QR.X KIS6QK4001, and the DFH/UFA CDFA-05-06.
\end{acknowledgments}

%

\end{document}


\maketitle

\newpage

\section{Magnetic field dependence of the electron tunneling rate}

The tunneling rate $\gamma_{In}$ for electrons from the charge reservoir into the quantum dot (QD) was determined by time-resolved resonant fluorescence (RF) n-shot measurements for every magnetic field; as the tunneling rate depends on the magnetic field.\cite{Amasha2008} 

\begin{figure*}[b!]
	\centering
	\includegraphics{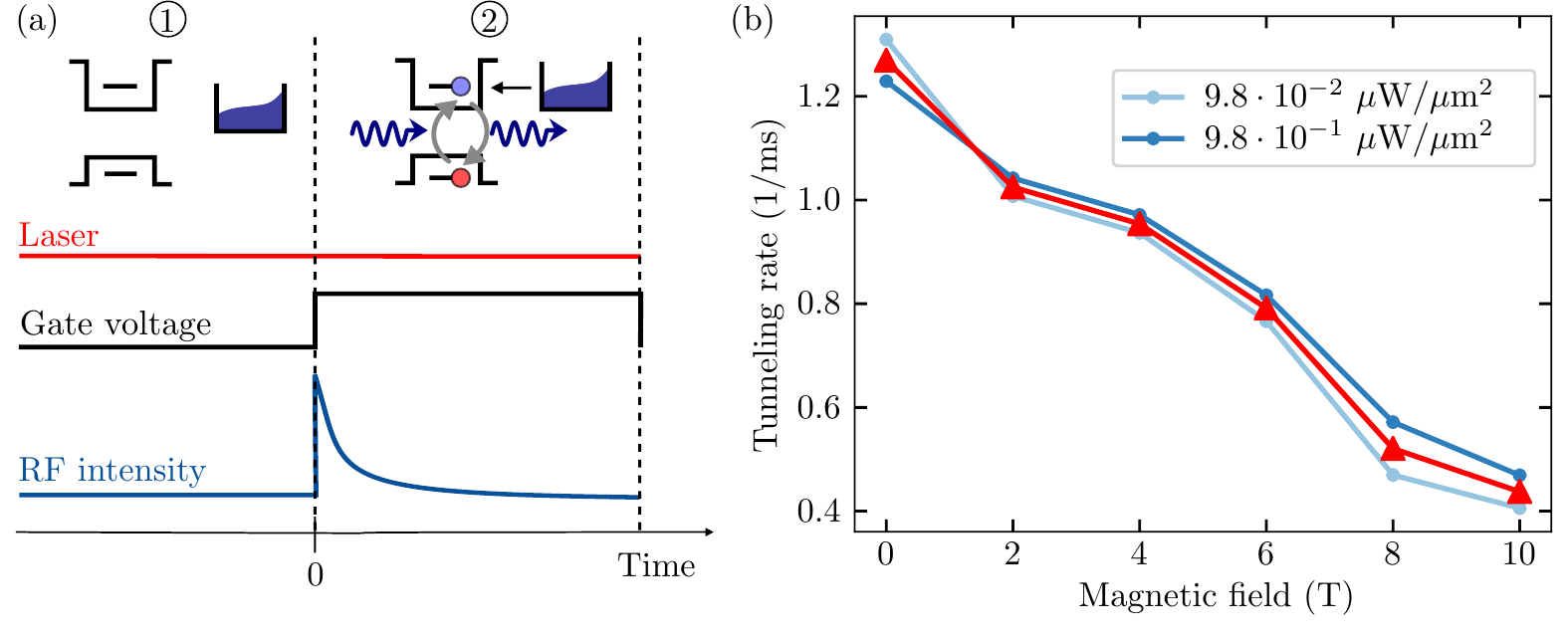}
	\caption{Time-resolved RF $n$-shot measurement to determine the tunneling rate of the electrons. (a) One shot consists of the preparation (1) and the probing (2) of the QD. (1) The QD is prepared empty at $V_\mathrm{G}=0\,$V. The laser is switched on, however, the exciton transition is not in resonance with the laser frequency. (2) The gate voltage is switched to $V_\mathrm{G}=0.5\,$V, where the exciton transition is now in resonance and RF intensity is detected until an electron tunnels into the QD. An exponential decay is observed after accumulation of many cycles.  (b) Tunneling rate of the electron into the QD measured for two laser intensities (blue). The tunneling rate does not depend on the laser intensity and the mean value (red) was used for the model in the main text.}
	\label{fig:Messung}
\end{figure*}

The pulse scheme for the n-shot measurements is shown in Fig.~\ref{fig:Messung}(a). The excitation laser (red line in Fig.~\ref{fig:Messung}(a)) is constantly switched on. Its frequency is set to $325.759\,$THz, so that the exciton transition is in resonance at a gate voltage of $0.5\,$V. The exciton transition is forbidden in equilibrium, as the dot is charged with one electron at this gate voltage. The QD is first emptied by an applied gate voltage of $V_g=0\,$V, where the Fermi energy in the reservoir is below the s-shell of the dot (black line in Fig.~\ref{fig:Messung}(a)). At $t=0$, the gate voltage is switched back to $0.5\,$V, where the exciton transition is in resonance with the laser frequency until an electron has tunneled into the QD and the RF intensity quenches. In an $n$-shot measurement, the RF intensity shows a decreasing transient (see blue line in Fig.~\ref{fig:Messung}(a)) with an exponential decay that corresponds to the tunneling rate $\gamma_{In}$. Figure \ref{fig:Messung}(b) displays the tunneling rate versus applied magnetic field for excitation laser intensities between $9.8\cdot10^{-2}\,\mu \mathrm{W}/\mu \mathrm{m}^2$ (bright blue) and $9.8\cdot10^{-1}\,\mu W/\mu m^2$ (dark blue). In red, the mean value for the tunneling rate for every magnetic field is depicted. It decreases with increasing magnetic field from $(1.27\pm0.04)\,\mathrm{ms}^{-1}$ at $0\,$T down to $(0.44\pm0.03)\,\mathrm{ms}^{-1}$ at 10\,T.

\section{Occupation probability of the trion state}

\begin{figure*}[htpb]
	\centering
	\includegraphics{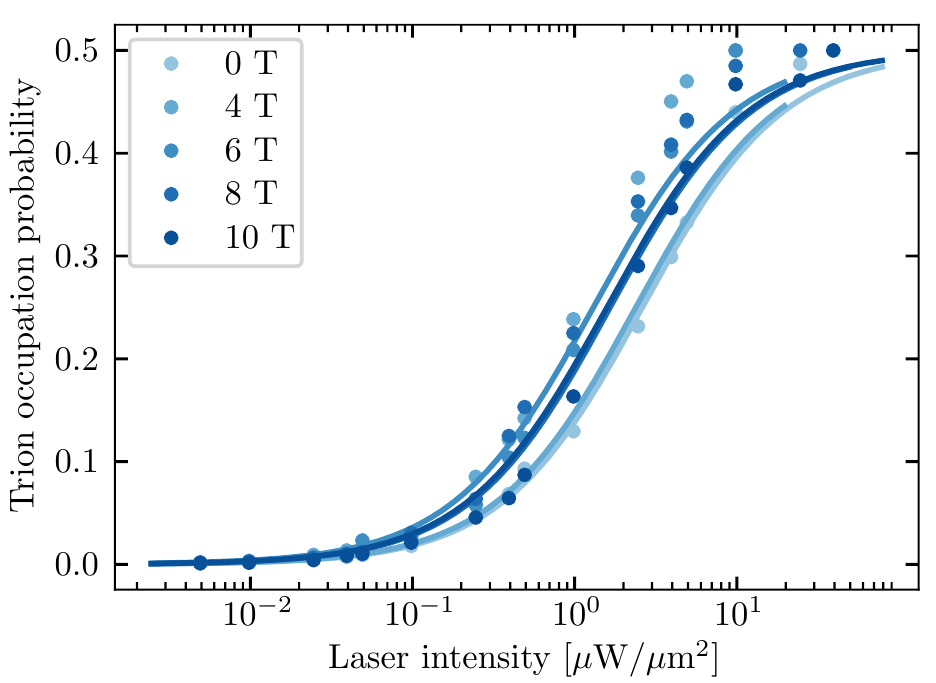}
	\caption{Occupation probability of the blue trion versus laser intensity for the different applied magnetic fields. The occupation probability is given by the resonance fluorescence (RF) counts of the blue trion transition at the beginning of the transient at $t=0$, see Fig.~3 in the main part of this paper. These RF intensities can be fitted with the function for the occupation probability of a two level system (blue lines).
	}
	\label{S2}
\end{figure*}

If the QD is prepared in the spin-up state $\ket{\uparrow}$ in a magnetic field $B$, the blue trion $\ket{\uparrow\downarrow\Uparrow}$ can resonantly be excited and RF can be measured. With increasing intensity of the resonant laser, the occupation probability of the QD with a trion $n$ increases until it saturates at 50\,\%, when at high laser intensities the QD is occupied with a trion half of the time.\cite{Loudon2000} Assuming only radiative recombination, the measured RF intensity is proportional to $n$ and can be used to determine the occupation probability. However, the Auger recombination and the spin-flip Raman process quench the intensity of the trion transition in equilibrium measurements. Thus, the trion occupation probability can only be measured in non-equilibrium before the Auger recombination and/or spin-flip Raman scattering has been occurred. This is the case for $t=0$ in the n-shot measurements in Fig.~3 in the main text, when the resonant laser is switched on and the QD is still occupied with one electron in the s-shell. For each magnetic field, a set of n-shot measurements for laser intensities between $4.9\cdot 10^{-3}\,\mu $W$/\mu $m$^2$ and $9.8\,\mu $W$/\mu $m$^2$ was measured. 
As discussed in the main part, at $t=0$ the quantum dot is in thermal equilibrium that determines the occupation probability of the spin-up $\ket{\uparrow}$ and spin-down state $\ket{\downarrow}$. As we only excited on the blue trion transition, Fig.~\ref{fig:S2} shows the blue trion $\ket{\uparrow\downarrow\Uparrow}$ occupation probability at the beginning of the transient at $t=0$. 
The ratio between spin-up/spin-down state at $t=0$ only depends on the magnetic field and not on the laser intensity and, hence, the  probabilities remain the same within one set of measurements at fixed magnetic field. Therefore, changing the magnetic field changes the entire curve in Fig.~\ref{S2} just by a different scaling factor and the data can still be fitted by the steady state occupation probability of a two level system\cite{Loudon2000} (blue lines). This trion occupation probability is used as a fixed parameter $n$ in the differential equation in the main part of this paper.

\section{Discussion of the fitting procedure}

\begin{figure*}[htpb]
	\centering
	\includegraphics[width=\textwidth]{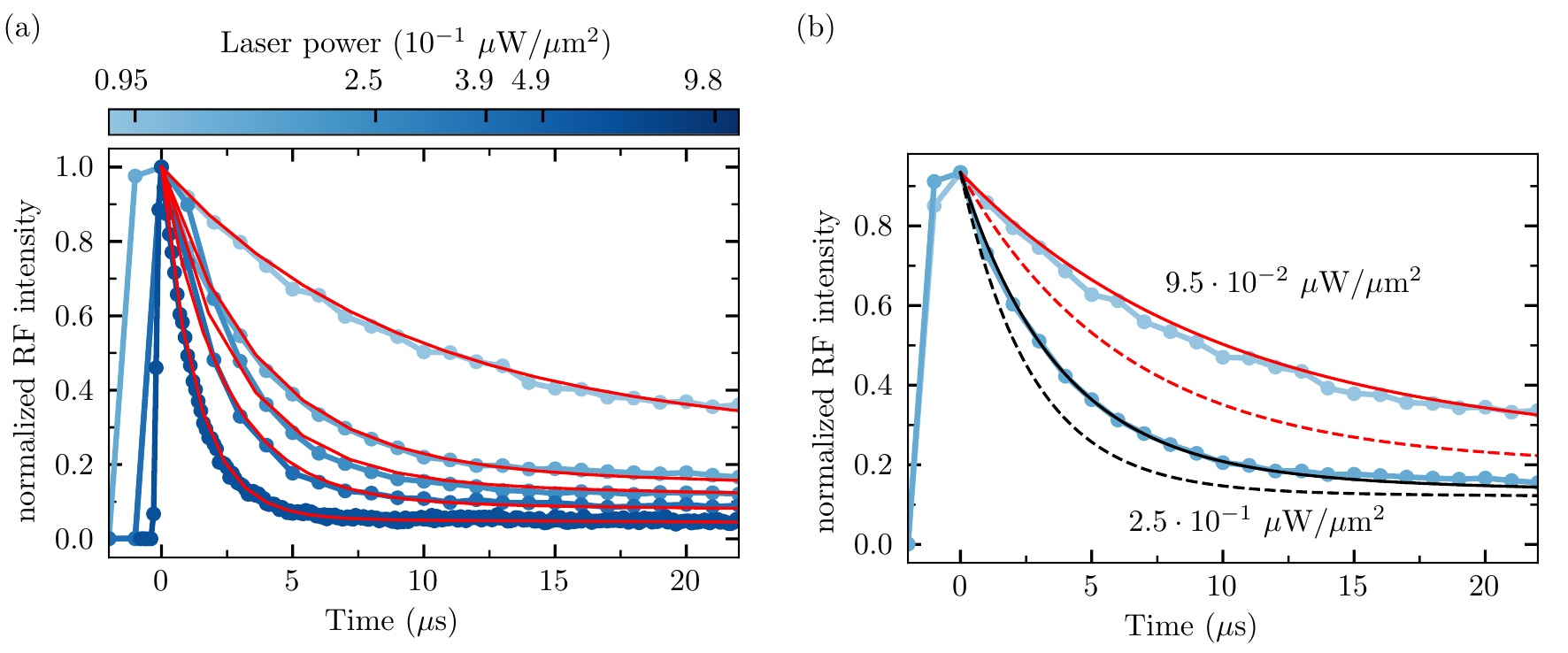}
	\caption{(a) The transients are normalized to unity in this graph, i.~e.~ the thermal distribution of spin-up to spin-down state in equilibrium is neglected. The red solid lines are fits to the normalized transients that are still in very good agreement with the data points. (b) The solid lines show the best fits to the transients. The dashed lines display two fits where the Auger rate was changed by 50\% and kept fixed in a second fitting procedure. This demonstrates that the spin-flip Raman rate cannot compensate for an off Auger rate.}
	\label{S3}
\end{figure*}

The transients in Fig.~3 in the main part are normalized to the occupation probability of the spin-up state in thermal equilibrium. We measured the exciton splitting in a magnetic field (the g-factor). We neglected the splitting of the hole spin states to obtain the electron spin splitting and afterwards, for instance, the occupation probability of the spin-up of about 65~\% at a magnetic field of $B= 4\,$T. The transients in Fig.~3 were normalized to this value and we have checked that this assumption has only a minor effect on the fitting procedure and the rates obtained. Figure \ref{S3}(a) shows the transients for an applied magnetic field of $B=10\,$T, where we neglected the occupation probability by the thermal distribution and started with the assumption of 100\% of electrons are in the spin-up state $\ket{\uparrow}$. The fits still give a very good agreement with the data and the rates are only slightly changed within an error margin of about 5 \%.

Moreover, we checked that we cannot exchange the Auger $\gamma_A$ and the spin-flip Raman rate $\gamma_R$ to obtain the same agreement in the fitting procedure with the experimental data. The dashed lines in Figure \ref{S3}(b) shows fits to the transients after the Auger rate has been changed by 50\% while the spin-flip Raman rate was still a free fitting parameter. Clearly visible is the strong deviation from the best fit to the data, shown as solid red and black line, i.~e.~the spin-flip Raman rate can not compensate for an off Auger rate in the fitting procedure.  

\begin{figure*}[htpb]
	\centering
	\includegraphics{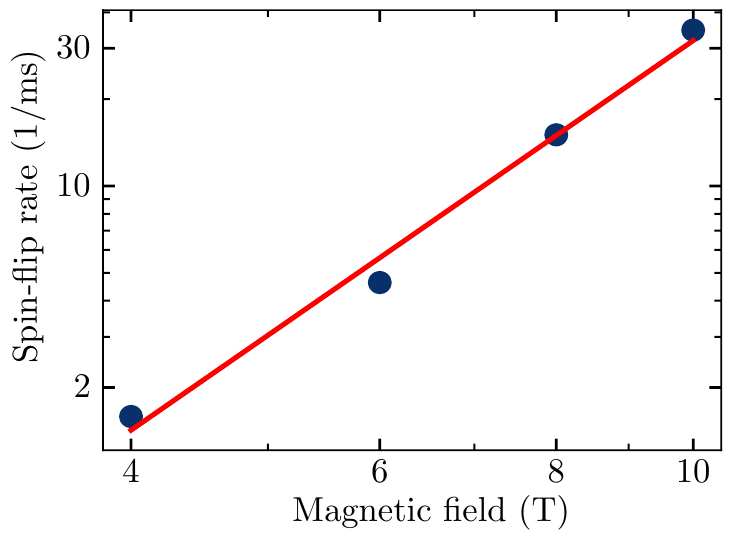}
	\caption{Double-logarithmic plot of the spin-flip rate versus the applied magnetic field. The magnetic field dependence of the spin-flip rate follows a power law (red) with an exponent of $3.4$.
	}
	\label{S4}
\end{figure*}

Finally, we compared the spin-flip rate and its dependence on the applied magnetic field with the literature, where a power law dependence with exponents between 4 and $5.3$ was obtained in Kroutvar et al. \cite{Kroutvar2004,Lu2010} and an exponent of about 5 in Lu et al.\cite{Lu2010}. Figure \ref{S4} shows our spin-flip rate on a double-logarithmic scale versus the applied magnetic field. A linear regression to the data gives an exponent for the power law dependence of 3.4, in agreement with the literature. This finally demonstrates the reasonability of our time-resolved measurements in combination with a fitting procedure that is based on a rate equation model. 

\newpage

\providecommand{\latin}[1]{#1}
\makeatletter
\providecommand{\doi}
{\begingroup\let\do\@makeother\dospecials
	\catcode`\{=1 \catcode`\}=2 \doi@aux}
\providecommand{\doi@aux}[1]{\endgroup\texttt{#1}}
\makeatother
\providecommand*\mcitethebibliography{\thebibliography}
\csname @ifundefined\endcsname{endmcitethebibliography}
{\let\endmcitethebibliography\endthebibliography}{}